\documentclass{article}
\usepackage{spconf, amsmath, graphicx, tabularx, amssymb}

\usepackage{enumitem}
\setlist{nosep, leftmargin=14pt}

\usepackage{mwe} % to get dummy images

\title{3D GRID-ATTENTION NETWORKS FOR INTERPRETABLE AGE AND ALZHEIMER'S DISEASE PREDICTION FROM STRUCTURAL MRI}
\name{Pradeep Lam, Alyssa H. Zhu, Iyad Ba Gari, Neda Jahanshad, Paul M. Thompson}
\address{Imaging Genetics Center, Mark and Mary Stevens Neuroimaging and Informatics Institute, \\ Keck School of Medicine, University of Southern California, Los Angeles, CA, USA}
\begin{document}
%\ninept
%
\maketitle
\begin{abstract}
We propose an interpretable 3D Grid-Attention deep neural network that can accurately predict a person's age and whether they have Alzheimer's disease (AD) from a structural brain MRI scan. Building on a 3D convolutional neural network, we added two attention modules at different layers of abstraction, so that features learned are spatially related to the global features for the task. The attention layers allow the network to focus on brain regions relevant to the task, while masking out irrelevant or noisy regions.  In evaluations based on 4,561 3-Tesla T1-weighted MRI scans from 4 phases of the Alzheimer's Disease Neuroimaging Initiative (ADNI), salience maps for age and AD prediction partially overlapped, but lower-level features overlapped more than higher-level features. The brain age prediction network also distinguished AD and healthy control groups better than another state-of-the-art method. The resulting visual analyses can distinguish interpretable feature patterns that are important for predicting clinical diagnosis. Future work is needed to test performance across scanners and populations. 
\end{abstract}

\section{Introduction}
\label{sec:intro}

BrainAge \cite{Katja} estimation refers to the task of estimating a person’s age from their brain imaging data; this task is often performed 
by computing features from healthy individuals’ structural MRI data (which may be supplemented by diffusion or functional MRI \cite{Smith1}). 
A statistical or machine learning model is then fitted to predict age, and tested on previously unseen subjects’ data. The ‘gap’ or ‘delta’ 
between a person’s estimated BrainAge and their true chronological age has been used as a marker of accelerated or delayed brain aging.
BrainAge gap has shown to be an accurate predictor of conversion from mild cognitive impairment (MCI) to Alzheimer's disease \cite{Gaser}.

Successful BrainAge prediction algorithms have been developed based on linear regression, using SVD to reduce the input feature dimension \cite{Smith1}, 
and classical machine learning methods such as support vector regression \cite{Han}. More recently, deep learning methods such as convolutional neural networks (CNN) 
have been applied to preprocessed and raw T1-weighted MRI data \cite{Cole}, and adapted to accommodate multiple datasets using transfer learning \cite{Jonsson}.

However, these networks can overfit to the task of age prediction and be invariant to disease \cite{Bashyam}, limiting the utility of the derived
brain age as a biomarker. Additionally, \cite{Buttler} brings up concerns regarding correcting for true age when predicting brain 
age delta and stresses that more interpretable methods need to be developed.

We developed a model that can learn BrainAge from distributed patterns in the image at different levels, maintaining sensitivity to disease without limiting training 
capacity, as in \cite{Bashyam}. Our model also uses attention to produce saliency maps for features at different levels of complexity.

Attention layers allow the network to focus on regions relevant to the task while masking out noise or irrelevant regions.
They have been used in medical imaging analysis and have improved accuracy in classification \cite{Schlemper}  and segmentation \cite{Sinah} tasks.
Our model uses a 3D version of grid-attention  originally proposed for fetal ultrasound screening by \cite{Schlemper}.
Our contributions are as follows:
\begin{itemize}
    \item We propose a new architecture for age prediction and Alzheimer's disease classification based on structural MRI.
    \item Our method shows group differences between brain age delta for Alzheimer's disease and healthy controls without limiting the accuracy of age prediction.
    \item Our method can visualize salient regions for classification at different points in the model with different complexities of features. 
    \item We compare the saliency maps for age with those important for Alzheimer's classification and find that lower-level features have significantly more overlap 
    than higher level features. To our knowledge we are the first to introduce this visual analysis to distinguish interpretable feature patterns used to predict age versus Alzheimer's disease classification.
\end{itemize}

\vspace{-1mm} %%HACK
\begin{figure}[htb]
    \centering
    \includegraphics[width=\linewidth]{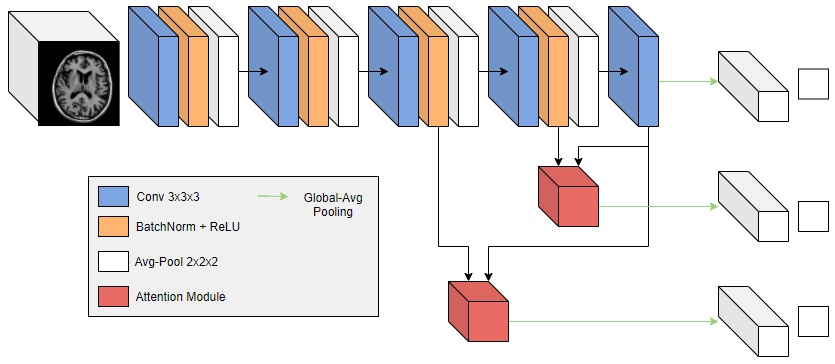}
    % \vspace{-5mm}  %%HACK
    \caption{Model Architecture}
    \label{fig:model_architecture}
\end{figure}
\vspace{-1mm}  %%HACK

\section{METHOD}
\label{sec:format}

\subsection{3D Grid-Attention Main Network}
\label{ssec:subhead}
The backbone of our model is a simple 3D convolutional neural network (Figure 1).
Esmaeilzadeh et al.\cite{Esmaeilzadeh} show that simple convolution models tend to  
outperform more complex models for classification of Alzheimer's disease by 
reducing overfitting.

The network consists of 4 convolution blocks followed by a final convolutional layer.
Each convolution block consists of a single 3D convolution layer (kernel size=3)
followed by a batch norm and ReLU activation and an average-pooling layer (kernel size=2, stride=2).
We find that average-pooling performs slightly better than max pooling for disease 
classification and our results are consistent with those of Jin et al. \cite{Jin}.
Global-average-pooling (GAP) is performed on the output of the final convolutional layer
and the resulting feature vector is input to a dense layer with varying number of nodes
depending on the classification/regression task.
The numbers of filters for each convolution are as follows: [32, 64, 64, 128, 128].
The number of output nodes is 23 for the age prediction task and 2 for Alzheimer's disease
classification.

\vspace{-1mm} %%HACK
\begin{figure}[htb]
    \centering
    \includegraphics[width=\linewidth]{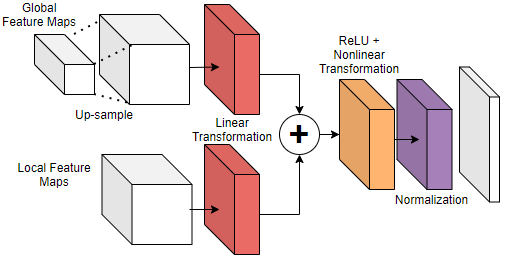}
    % \vspace{-5mm}  %%HACK
    \caption{Grid-Attention Mechanism}
    \label{fig:attention_mechanism}
\end{figure}
\vspace{-1mm}  %%HACK

\subsection{Grid-Attention Mechanism}
\label{ssec:subhead}
We use two attention modules based on \cite{Schlemper} to better integrate low-level features into the age prediction and AD classification.
The attention module is designed so that features learnt are spatially related to the global features for the task.
The architecture of our attention module is shown in Figure 2. The "attention score" is represented by the following equation:
\begin{equation}
\alpha_{\mathit{l}}= \pmb{\psi}\sigma\left (\mathbf{W}_{\mathit{f}}\mathbf{f}_{\mathit{l}}+\mathbf{W}_{\mathit{g}}\mathbf{g}\right ) + \mathbf{b}_{\psi}
\end{equation}
where \(\mathbf{f}_{\mathit{l}}\in \mathbb{R}^{ H_{\mathit{l}} \times W_{\mathit{l}} \times D_{\mathit{l}}}\) represents the output activation maps for block \(\mathit{l}\)
before the pooling operation and \(\mathbf{g}\in \mathbb{R}^{H_{\mathit{g}} \times W_{\mathit{g}} \times D_{\mathit{g}}}\) represents the output activation maps of the final 
convolution layer before global average pooling. 

\(\mathbf{W}_{\mathit{f}}\in \mathbb{R}^{\mathit{C_{int}\times}C_{f}}\), \(\mathbf{W}_{\mathit{g}}\in \mathbb{R}^{\mathit{C_{int}\times}C_{g}}\), and
\(\mathbf{W}_{\mathit{\psi}}\in \mathbb{R}^{\mathit{C_{int}\times}1}\) are linear transformations, \(\sigma\) is a non-linear transformation, and 
\(\mathbf{b}_{\mathit{\psi}}\in \mathbb{R}\) is a bias term. We choose ReLU for the non-linearity and find that it results in more accurate predictions than Softmax.

The resulting attention score \(\alpha_{\mathit{l}}\) has a dimensionality of 1xHxWxD. This score is normalized per module by subtracting \(min(\alpha_{\mathit{l,i,j,k}})\)
and then dividing by \(\sum{\alpha_{\mathit{l,i,j,k}}}\). This normalized attention score is multiplied by each input activation map to produce the output of the module. Finally, 
a sum is taken over feature maps and is connected to a dense layer with the same number of nodes as the backbone network. Thus, regions with a higher normalized attention score are more 
relevant to the task and are highlighted, while noise and irrelevant features are masked out.

Our implementation connects attention modules after convolutions 3 and 4, as shown in Figure 1. We implement \(\mathbf{W}_{\mathit{f}}\) and \(\mathbf{W}_{\mathit{g}}\) as 1x1x1 convolution operations and 
use n=64, 128 intermediate filter dimensions respectively. Global feature maps are upsampled using trilinear upsampling to match the dimensionality of earlier activation maps. We aggregate the results by calculating the average of the dense output layers for the two attention modules as well as the backbone network.

\subsection{Kullback-Leibler Divergence (KLD) Loss}
\label{ssec:subhead}
We formulate age prediction tasks similarly to \cite{Peng} and \cite{Lam}. Age is represented by a discrete Gaussian distribution with mean equal to the true age of 
a subject and with a standard-deviation of 2. We train our age prediction model to minimize the Kullback-Leibler Divergence (KLD) between this 
distribution and the soft-max network output. There are 23 bins in our distribution: each bin represents an interval of 2 years and corresponds to 23 nodes in the age prediction network.

\subsection{Data and Preprocessing}
\label{ssec:subhead}
Our experiments analyze brain structural (3T) T1-weighted MRI scans from the Alzheimer's Disease Neuroimaging Initiative (ADNI) dataset. We use scans from ADNI1, ADNI2, ADNI3, and ADNI-GO studies; our dataset consists of 4,561 scans 
in total (2,004 non-accelerated, 2,557 accelerated) and contains 329 unique subjects with AD (191 male, 138 female, mean age=75.7 years, standard deviation=8.18) and 687 unique healthy controls (288 male, 399 female, mean age=73.0, std=7.0).
Scans are processed through a standard processing pipeline consisting of reorientation using \cite{FSL}, skull stripping using HDBet \cite{Isensee}, nonparametric intensity normalization for bias field correction using N4BiasCorrection \cite{N4Bias}, 
and 6-dof linear registration to MNI152 space using FSL \cite{FSL}. Scans are subsequently down-sampled to 2-mm resolution using ANTS \cite{ANTS}. The final dimensions of each scan are 91x109x109.

\subsection{Experiment Set-up and Training}
\label{ssec:subhead}
We split our data into 3 non-overlapping sets: training, validation (for hyperparameter tuning and model stopping), and test sets. We create an age-balanced validation set of n=109 (mean age=74.76 years, std=8.2, 53 AD, 56 CN) and use the remaining scans to create 10 pairs of training/testing sets (10-fold split). As our dataset may contain multiple scans for a single subject, we split our 
data by unique subject ID to ensure that no subjects are in common between training, validation, and testing sets within a single fold. Follow-up and duplicate scans are used for data augmentation for
our training set but are filtered out from our validation and testing sets. For our age-prediction models, we further filter out subjects with Alzheimer's disease from our training  
and validation sets. For testing, we use a single scan per subject chosen at random; for standardization, we use a non-accelerated scan if it exists but include accelerated scans for subjects without any non-accelerated scans. \cite{Vemuri} shows comparable group discrimination in accelerated vs. non-accelerated scans. For training our age models, we sample
from a unique set of subject IDs and randomly select at most one of the multiple scans for each subject. For training the Alzheimer's classification model, we split our subjects into two sets based on diagnostic class and sample equally from both sets. 

Scan intensities are standardized to mean zero and unit variance for non-zero (non-background) pixels.  We augment our training data using shift (uniform (-0.1, 0.1)*std) and scale (uniform (-0.1, 0.1)) intensity augmentation on non-zero pixels and random (p=0.5) hemispherical mirroring. 

We initialize our model weights using Kaiming weight initialization \cite{He} and train our model using the ADAM \cite{Kingma} optimizer, with batch size of 16 and learning rate of 1e-5. We use a weight decay of 1e-2 to prevent overfitting. We minimize KLD Loss (detailed in section 2.3) to train our age model and 2-class cross entropy for the Alzheimer's disease classification model and train our models until 10 epochs without performance improvement on the validation set. Test data is evaluated on the state with the lowest validation loss.

\section{RESULTS}
\label{sec:pagestyle}

\subsection{Models Results}
\label{ssec:subhead}
Our final age prediction model has a mean absolute error (MAE) of 3.96 years on healthy controls (RMSE=5.17, Corr=0.686) for n=991 subjects from the ADNI dataset. For comparison, we replicated our training/evaluation 
procedure with \cite{Peng} which received state-of-the-art performance on the UK Biobank dataset and received an MAE of 4.36 on healthy controls (RMSE of 5.56, Corr=0.66). Although state-of-the art shows much lower MAE for brain age prediction (2.86 MAE \cite{Lam}), we attribute this to having far more training samples (7,312 scans). Our final AD prediction model has an accuracy of 0.868 (balanced accuracy=0.84, precision=0.81, recall=0.76) over n=991 subjects.

\vspace{-1mm} %%HACK
\begin{figure}[htb]
    \centering
    \includegraphics[width=\linewidth]{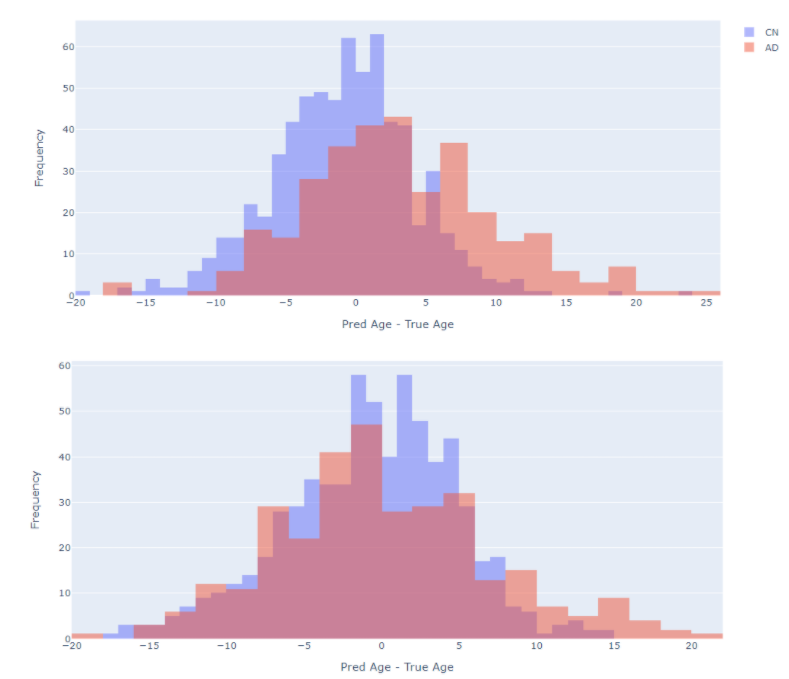}
    % \vspace{-5mm}  %%HACK
    \caption{The histograms show the brain age delta for our attention model (top) and for an alternative state-of-the-art model \cite{Peng} (bottom).
    Our method shows a greater sensitivity to AD: the average brain age delta is 3.09 (SD=6.76) for subjects with AD (Orange) and -0.89 (SD=5.10) for healthy controls(Blue). 
    The compared model has an average brain age delta of 1.00 (SD=6.99) for subjects with AD and -0.58 (SD=5.62) for healthy controls.}
    \label{fig:pad_hist}
\end{figure}
\vspace{-1mm}  %%HACK

\subsection{Sensitivity to Disease Effects}
\label{ssec:subhead}
The distributions of brain age delta (the difference between predicted and true age) for AD and healthy controls are shown in Figure 2 (top). The MAE for AD subjects is 5.7 years (RMSE: 7.4 years). 
For comparison, we also included the distribution for the model based on \cite{Peng} (bottom); the MAE for AD subjects is 5.56 years (RMSE: 7.01 years).
As can be observed, our model detects a greater group effect on BrainAge for AD. This observation corresponds to a higher Cohen's d effect size (95 percent confidence interval) for our model (0.66 $\pm$ 0.13; P-value $<$ 0.00001) as compared to [13] (0.25 $\pm$ 0.13; P-value = 0.0002). We attribute this to our model using fewer convolutional 
blocks for the backbone architecture and the use of attention modules to enforce the use of lower level features for age prediction.

\vspace{-1mm} %%HACK
\begin{figure}[htb]
    \centering
    \includegraphics[width=\linewidth]{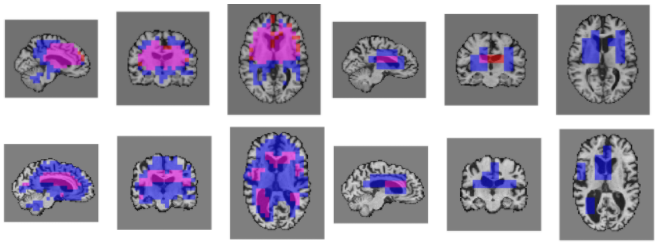}
    % \vspace{-5mm}  %%HACK
    \caption{Saliency visualization maps produced by the model for two different subjects. Three left panels show saliency maps for the first attention module 
    and the rightmost three panels correspond to the second module. Red colors denote voxels relevant for AD classification, blue voxels for age prediction, and
    purple colors show the region of overlap of voxels relevant for each task. Top row uses a threshold of 2 standard deviations above mean attention,
    while the bottom row uses a threshold of 1 std above the mean.
    }
    \label{fig:att_maps}
\end{figure}
\vspace{-1mm}  %%HACK

\subsection{Visualization and Comparison}
\label{ssec:subhead}

\begin{table}[h!]
\centering
\begin{tabularx}{0.45\textwidth} { 
  | >{\centering\arraybackslash}X 
  | >{\centering\arraybackslash}X 
  | >{\centering\arraybackslash}X | }
 \hline
  & A1 (CN/AD) & A2 (CN/AD) \\
 \hline
 \(\mu +2\sigma\)  & 0.39/0.39 & 0.29/0.30 \\ 
\hline
 \(\mu +1\sigma\)  & 0.43/0.41  & 0.20/0.26  \\
\hline
\(\mu +0.5\sigma\) & 0.52/0.52 & 0.10/0.20 \\ 
\hline
\end{tabularx}
\caption{Average overlap scores between saliency maps for the age prediction and AD prediction tasks, separate for AD subjects (AD) versus healthy controls (CN)
for varying intensity levels of thresholded attention value. Column A1 is the average of scores for the first attention module while column A2 
is the average of scores for the second attention module.}
\label{table:1}
\end{table}

We visualize region saliency using the attention masks (normalized attention scores) produced by the attention modules. We upsample these masks to the original size of our 
scan using nearest neighbor interpolation and mask out artifacts by multiplying by the binary brain mask. For analysis, we apply binary thresholding based on the 
mean attention value. Figure 4 shows examples of saliency maps produced by our model at different thresholds. We note that saliency maps produced by the network differ
by subject and are based on relevant areas used to classify each scan.

In addition, we analyze the difference in region overlap between salient features for age prediction and AD classification. For each subject, we calculate the 
Jaccard score between the thresholded saliency maps for AD classification and age prediction:
\begin{equation}
\mathbf{J} = \frac{|AD \cap AGE|}{|AD \cup AGE|}
\end{equation}
where AD and AGE are binary (thresholded) importance regions. Results are presented in Table 1. As the \(\sigma\) threshold decreases, there is a 
greater difference in overlap between attention modules. For \(\sigma=1\) there is a 0.52 overlap for the first attention 
module compared to only 0.1 for the subsequent module. This implies that feature complexity plays 
a crucial role in common region saliency between tasks. Visually, this can also be seen in Figure 4. Age prediction focuses on a 
much larger region of features compared to AD classification which focuses mainly on the ventricles. This may explain why more complex age prediction models 
tend to more accurately predict age in diseased subjects. We also observed little difference between average scores for subjects with
AD versus healthy controls.

\section{CONCLUSION}
\label{sec:typestyle}
In this work, we propose an interpretable 3D Grid-Attention network that can accurately predict age and Alzheimer's disease from structural MRI. We compare saliency maps 
produced by this model between age prediction and Alzheimer's disease and find that the features used for each task do overlap, but are partially distinct: age prediction tends to use more widespread features, while AD prediction makes use of ventricular expansion, which is mild in normal aging but can be severe in AD. When brain aging and AD classification models are forced to use the same small set of features, it is almost inevitable that people with AD appear to have accelerated aging. The current brain aging model also focuses on brain regions that are less affected by pathology. This may also explain why complex deep learning models trained for brain age prediction may be somewhat invariant to detecting shifts in brain age, in disease \cite{Bashyam}. The overlap of the salience maps for each task offers a more intuitive understanding of what image regions are used for each task, and their differences encourages further study of localized brain aging for biomarker development. Future work will involve extending our analysis to include 
subjects with MCI, a precursor to Alzheimer's disease and extending our architecture to predict both age and Alzheimer's disease status simultaneously using multi-task learning. Further work should also test the approach on multiple datasets to fully assess generalization performance for other scanners and populations. Ongoing work in domain adaptation and adversarial networks will also be needed to train models that generalize well and offer invariance to scanning protocols \cite{Guan} \cite{Dinsdale}.

\section{Compliance with Ethical Standards}
\label{sec:ethics}

Data used in this article were obtained from the public Alzheimer's Disease Neuroimaging Initiative (ADNI) database (adni.loni.usc.edu). Further ethical approval was not required, as we studied previously collected, anonymized, de-identified data.

\section{Acknowledgments}
\label{sec:acknowledgments}

This research was supported in part by the National Institutes of Health (NIH) under grants U01AG068057, R01AG059874, and RF1AG051710, and by a research grant from Biogen, Inc.

\bibliographystyle{IEEEbib}
\bibliography{refs}

\end{document}